\documentclass[%
 reprint,
superscriptaddress,
 amsmath,amssymb,
prl,
]{revtex4-2}

\usepackage{graphicx}
\usepackage{dcolumn}
\usepackage{bm}
\usepackage{color}
\usepackage{xspace}
\usepackage{multirow}
\usepackage{hyperref}
\usepackage{ulem}
\usepackage{here}

\definecolor{green}{rgb}{0,0.6,0.1}

\begin{document}

\preprint{APS/123-QED}

\title{Strong-coupling high-$T_{\rm c}$ superconductivity in doped correlated band insulators}

\author{Yusuke Nomura}
\email{yusuke.nomura@tohoku.ac.jp}
\affiliation{
Institute for Materials Research (IMR), Tohoku University, 
2-1-1 Katahira, Aoba-ku, Sendai 980-8577, Japan
}

\author{Motoharu Kitatani}
\affiliation{
Department of Material Science, University of Hyogo, Ako, Hyogo 678-1297, Japan
}

\author{Shiro Sakai}
\affiliation{
Physics Division, Sophia University, Chiyoda-ku, Tokyo 102-8554, Japan
}
\affiliation{
RIKEN Center for Emergent Matter Science, 2-1 Hirosawa, Wako, Saitama 351-0198, Japan
}

\author{Ryotaro Arita}
\affiliation{
RIKEN Center for Emergent Matter Science, 2-1 Hirosawa, Wako, Saitama 351-0198, Japan
}
\affiliation{
Department of Physics, The University of Tokyo,7-3-1 Hongo, Bunkyo-ku, Tokyo 113-0033
}

\date{\today}

\begin{abstract}
We explore the superconducting properties of the bilayer Hubbard model, which exhibits a high transition temperature ($T_{\rm c}$) for an $s_{\pm}$ pairing, using a cluster extension of the dynamical mean-field theory.
Unlike the single-layer Hubbard model, where the $d$-wave superconductivity emerges by doping the Mott insulator, the parent state of the bilayer system is a correlated band insulator.
Above $T_{\rm c}$, slight hole (electron) doping introduces a striking dichotomy between electron and hole pockets: the electron (hole) pocket develops a pseudogap while the other becomes a nearly incipient band.
We reveal that the superconductivity is driven by kinetic (potential) energy gain in the underdoped (overdoped) region.
We also find a very short coherence length, for which we argue the relevance to multi-orbital physics.
Our study offers crucial insights into the superconductivity in the bilayer Hubbard model potentially relevant to
La$_3$Ni$_2$O$_7$.
\end{abstract}

\maketitle


\paragraph{Introduction.}
The problem of high-$T_{\rm c}$ superconductivity (SC) in strongly correlated electron systems is one of the central issues in condensed matter physics. The canonical model that represents high-$T_{\rm c}$ SC resulting from electron correlation effects has played a crucial role in elucidating the mechanisms of its occurrence~\cite{RevModPhys.84.1383}. The Hubbard model is one of the representative models, and in particular, the single-band Hubbard model on a square lattice has been extensively studied as a model representing high-$T_{\rm c}$ cuprates. To date, various approaches from weak to strong correlations have established the existence of an SC phase with $d$-wave symmetry near the Mott insulating phase~\cite{PhysRevLett.95.237001,PhysRevLett.110.216405,PhysRevB.90.115137,PhysRevB.98.205132,doi:10.1126/science.aal5304,doi:10.1126/science.adh7691}.

Meanwhile, the bilayer Hubbard model has been studied as another canonical model for achieving high-$T_{\rm c}$ SC. In 1992, focusing on bilayer cuprates, Bulut {\it et al.} showed that increasing interlayer hopping could change the SC gap symmetry from $d$-wave to $s_{\pm}$-wave~\cite{PhysRevB.45.5577}. Since that discovery, the potential for high-$T_{\rm c}$ SC within the bilayer Hubbard model has been explored through various approaches~\cite{RevModPhys.84.1383,PhysRevB.104.245109,PhysRevB.99.140504,SciRep.6.32078,PhysRevB.84.180513,PhysRevB.80.224524,PhysRevB.77.144527,PhysRevB.66.184508,PhysRevLett.74.2303,PhysRevB.51.15540,PhysRevB.50.4159,PhysRevB.50.13419,PhysRevResearch.2.023156}. This issue has recently gained renewed attention in connection with elucidating the mechanisms behind the newly discovered bilayer nickelate high-$T_{\rm c}$ superconductor, La$_3$Ni$_2$O$_7$~\cite{Nature.621.493}. In La$_3$Ni$_2$O$_7$, the low-energy states are composed of the $d_{x^2-y^2}$ and $d_{3z^2-r^2}$ orbitals~\cite{PhysRevLett.131.126001,PhysRevLett.131.206501}. 
The roles these two orbitals play in SC, and whether they lead to the $d$-wave SC, similar to that observed in cuprates, or the $s_{\pm}$-wave SC, similar to that proposed for iron-based superconductors, remains an open question~\cite{arxiv2306.07275,Chin.Phys.Lett.41.017402,npj.Quantum.Matt.9.61,PhysRevLett.132.146002,NatureCommun.15.2470,PhysRevB.109.L180502,PhysRevB.108.L201121,PhysRevLett.131.236002,PhysRevB.109.104508,PhysRevB.108.L140505,arxiv.2409.17861,arxiv.2501.05254,PhysRevLett.133.096002,arxiv.2501.10409,PhysRevB.108.214522,arxiv.2412.21019,PhysRevB.109.165154,PhysRevB.110.024514,arXiv.2502.08425}.
Among these studies, a scenario has been proposed where the $d_{3z^2-r^2}$ orbital plays the role of the active band, giving rise to high-$T_{\rm c}$ $s_{\pm}$-wave SC through a mechanism inherent in the bilayer Hubbard model~\cite{PhysRevB.95.214509,PhysRevLett.132.106002}.

Both the single-layer and bilayer Hubbard models share the common feature of realizing high-$T_{\rm c}$ SC near half filling, but there are crucial differences concerning the insulating phase at half filling~\cite{PhysRevB.75.193103,Lee_2024}. In the case of the single-layer Hubbard model, the insulating phase is the Mott insulator, whereas in the bilayer Hubbard model, it is a correlated band insulator. 
For the electron and hole bands, the energy scale from the bottom of the band to the Fermi level is small, competing with the size of the SC gap, suggesting the presence of BEC (Bose-Einstein condensation)-like strong-coupling SC. 
While the instability toward SC in the normal state  in the bilayer Hubbard model has been explored, 
more expensive calculations of the SC state itself have not been conducted so far, except for the filling dependence of the order parameter~\cite{PhysRevB.109.165154}. 
As a result, the nature of superconductivity remains poorly understood.


In this Letter, we analyzed the bilayer Hubbard model using the cellular dynamical mean field theory (cDMFT), one of the most reliable non-perturbative techniques~\cite{Kotliar_2001,RevModPhys.77.1027}. 
First, we calculate the condensation energy, showing that kinetic-energy-driven SC occurs near the correlated band insulating phase and that it crosses over to a potential-energy-driven one as the carrier doping increases.
The transition temperature reaches its maximum just before the crossover region from kinetic to potential energy-driven SC, amounting to about 10 \% of the in-plane transfer energy. 
Further analysis of the spectral function as a function of hole (electron) doping reveals that in the electron (hole) band, where the Fermi energy is small, a pseudogap opens above $T_{\rm c}$, while the hole (electron) band exhibits conventional metallic behavior.
We also evaluate the coherence length of SC, finding that near the correlated band insulating phase, the coherence length is only a few times the lattice constant, which is extremely short. If La$_3$Ni$_2$O$_7$ is described by the bilayer Hubbard model, this corresponds to $\sim 1$ nm, aligning with the fact that La$_3$Ni$_2$O$_7$ has a high critical field ($H_{\rm c2}$) of $\sim$ 100 T~\cite{Nature.621.493,arxiv.2501.14584}. 


\paragraph{Model and Methods.}
We consider the bilayer Hubbard model on the square lattice, whose Hamiltonian reads
\begin{eqnarray}
\! \! \! \! \! 
\mathcal{H} \! = \!  
- t  \! \! \!  \sum_{\langle i,j \rangle, \alpha , \sigma }  \! \! c^{\sigma \dagger}_{\alpha i} c^{\sigma}_{\alpha j} 
- t_\perp \! \! \! \sum_{i, \alpha\neq \alpha', \sigma} \! \!  c^{\sigma \dagger}_{\alpha i} c^{\sigma}_{\alpha' i}  +U \! \sum_{i,\alpha} n^{\uparrow}_{\alpha i}  n^{\downarrow}_{  \alpha i }, 
\end{eqnarray}
where $c_{\alpha i}^{\sigma \dagger}$ ($c^{\sigma}_{\alpha i}$) creates (annihilates) an electron with spin $\sigma$ at site $i$ on the $\alpha$-th layer ($\alpha=1,2$) and $n^{\sigma}_{\alpha i}\equiv c_{\alpha i}^{\sigma \dagger} c^{\sigma}_{\alpha i}$.
$t$ and $t_\perp$ are intralayer nearest-neighbor hopping and interlayer hopping, respectively. 
We set $t=1$ as the energy unit and study the case of $t_\perp = 2$ and $U = 8$.
In this study, we focus on hole doping, but the results for electron doping can be obtained straightforwardly due to the particle-hole symmetry of the model. 

We solve this model by cDMFT. 
We incorporate intralayer short-range correlations within the cluster size of $2 \! \times \! 2$ and the interlayer correlations.   
We employ the Nambu formalism to treat the SC phase.
This approach is complementary to that employed in Refs.~\cite{PhysRevB.104.245109,PhysRevB.99.140504,PhysRevB.84.180513}, where the pairing instability was investigated through calculations above $T_{\rm c}$ using the dynamical cluster approximation~\cite{RevModPhys.77.1027}.
For the impurity solver, we use the continuous-time quantum Monte Carlo method with an interaction expansion~\cite{Rubtsov_2004,Rubtsov_2005} developed in Ref.~\cite{Nomura_2014}, which implements the submatrix update algorithm~\cite{Gull_2011} to make the computation at low temperatures feasible~\footnote{An open-source CT-INT program using the submatrix update as in Ref.~\cite{Nomura_2014} is available in Ref.~\cite{Shinaoka_2020}.}.

If we take the bonding and antibonding basis, the noninteracting band dispersions $\varepsilon_{\gamma {\bm k}}$ for the bonding ($\gamma={\rm B}$) and antibonding ($\gamma={\rm A}$) orbitals are given by 
$\varepsilon_{{\rm B} {\bm k}}=-t_\perp -2t(\cos k_x + \cos k_y)$
and
$\varepsilon_{{\rm A} {\bm k}}=t_\perp -2t(\cos k_x + \cos k_y)$, respectively. 
In this basis, the normal-state Green's function becomes diagonal and is given by 
 $G_{\gamma {\bm k}}( i \omega_\nu )  = \bigl [ i \omega_\nu + \mu - \varepsilon_{\gamma {\bm k}}  - \Sigma_{\gamma {\bm k}} (i \omega_\nu ) \bigr]^{-1}$ with the Matsubara frequency $\omega_\nu = (2\nu + 1 ) \pi T$ ($T=1/\beta$ is the temperature), the chemical potential $\mu$, and the self-energy $\Sigma_{\gamma {\bm k}} (i \omega_\nu )$. 
In the SC phase, the anomalous part of the self-energy $S_{\gamma {\bm k}} (i\omega_\nu)$ becomes nonzero.

In the current framework, we can directly obtain the self-energy at momenta ${\bm k}=(0,0)$, $(\pi,0)$, $(0,\pi)$, $(\pi,\pi)$.
To infer the full momentum dependence, we employ the periodization scheme~\cite{Kotliar_2001}.
Specifically, we adopt the cumulant periodization method~\cite{Stanescu_2006}, which ensures rapid convergence of the periodized self-energy with increasing cluster size~\cite{Sakai_2012}.
Hereafter, results for momenta other than ${\bm k}=(0,0)$, $(\pi,0)$, $(0,\pi)$, $(\pi,\pi)$ are obtained through this periodization scheme.

\paragraph{Phase diagram.}

\begin{figure}[tb]
\vspace{0.0cm}
\begin{center}
\includegraphics[width=0.47\textwidth]{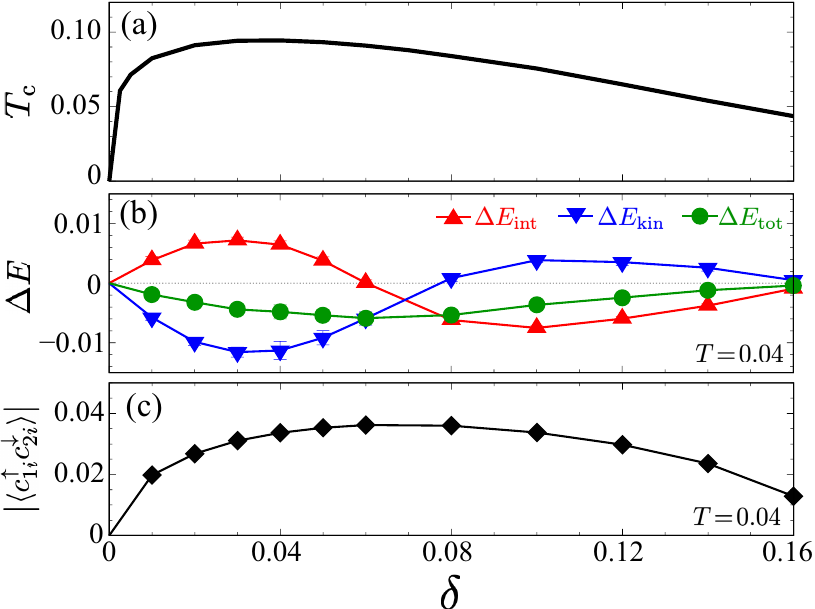}
\caption{Doping dependence of (a) $T_{\rm c}$,  (b) energy difference between SC and normal states at $T\!=\!0.04$, and (c) the SC order parameter $|\langle c_{1i}^\uparrow c_{2i}^\downarrow \rangle|$ at $T\!=\!0.04$.
}
\vspace{0cm}
\label{Fig1}
\end{center}
\end{figure}

\begin{figure*}[tb]
\vspace{0.0cm}
\begin{center}
\includegraphics[width=0.99\textwidth]{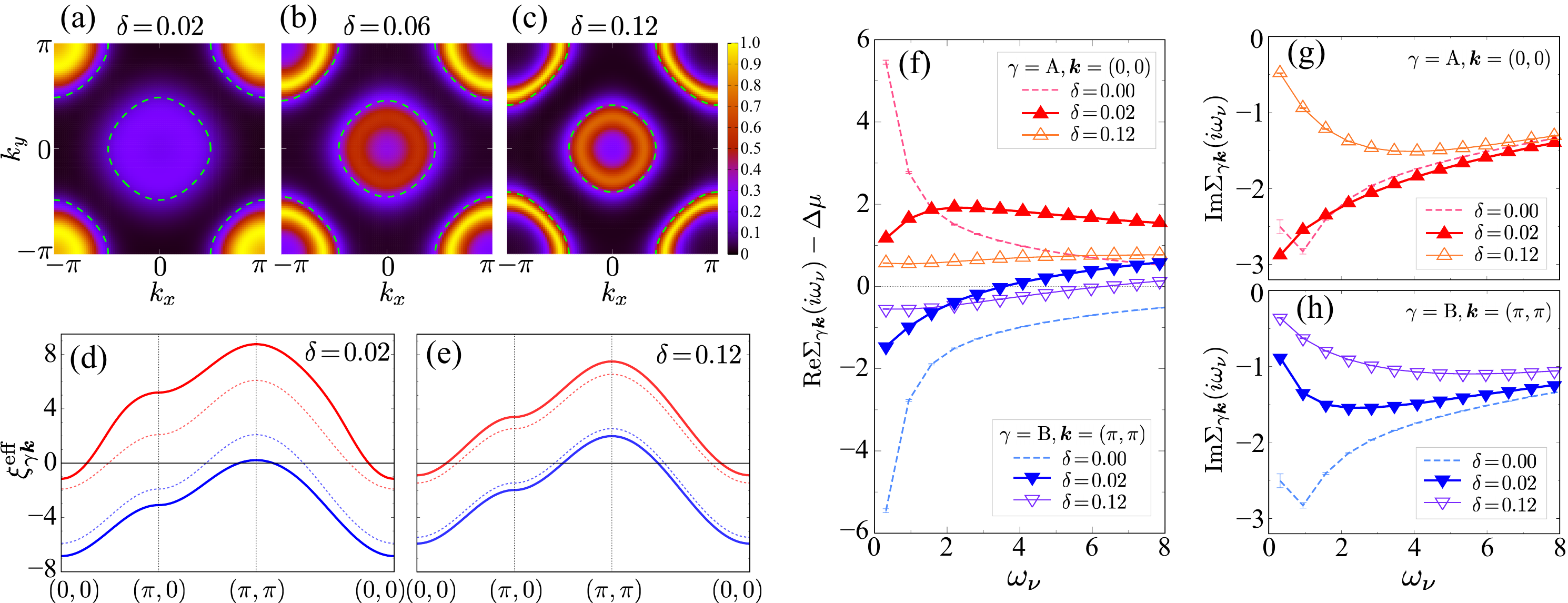}
\caption{
The normal state properties at $T=0.1$.
(a,b,c) The Fermi surface [$-\sum_\gamma  \beta G_{\gamma {\bm k}} (\tau \! = \! \beta/2)$ normalized by its maximum value].
The dashed curves show the Fermi surface at $U=0$. 
(d,e) The effective dispersion $\xi_{\gamma {\bm k}}^{\rm eff} = \varepsilon_{\gamma {\bm k}} + {\rm Re }\Sigma_{\gamma {\bm k}}(0) - \mu $ [where ${\rm Re }\Sigma_{\gamma {\bm k}}(0)$ is approximated by ${\rm Re }\Sigma_{\gamma {\bm k}}( i \omega_0)$]. 
The dotted curves show the dispersion at $U=0$.
(f) The real part of self-energy ${\rm Re}\Sigma_{\gamma {\bm k}}(i \omega_\nu) - \Delta \mu$ for antibonding and bonding bands at ${\bm k}=(0,0)$ and $(\pi,\pi)$, respectively.
$\Delta \mu$ is the change in the chemical potential between $U=8$ and $U=0$.
(g,h) The imaginary part of self-energy ${\rm Im }\Sigma_{\gamma {\bm k}}$ for antiboding band at ${\bm k}=(0,0)$ [panel (g)] and bonding band at ${\bm k}=(\pi,\pi)$ [panel (h)].
}
\vspace{0cm}
\label{Fig2}
\end{center}
\end{figure*}

Figure~\ref{Fig1}(a) illustrates the doping dependence of $T_{\rm c}$, revealing a clear dome-like shape. 
The maximum $T_{\rm c}$ reaches about 0.1, in consistency with previous studies with different methods~\cite{PhysRevB.66.184508,PhysRevB.84.180513}. 
This value is significantly higher than $T_{\rm c}^{\rm max} \! \simeq \! 0.05$ obtained from the $2\times2$ cDMFT calculation~\cite{Fratino_2016} for $d$-wave SC in the single-layer Hubbard model. 
Notably, in the underdoped regime, the SC is driven by kinetic energy gain, indicating a strong-coupling nature [Fig.~\ref{Fig1}(b)]. 
As doping increases, a crossover to a potential-energy-driven mechanism occurs. 
The SC order parameter also exhibits a dome-like shape [Fig.~\ref{Fig1}(c)].

\paragraph{Normal state properties.}
Now, we discuss how the SC emerges by investigating the normal state properties at $T=0.1$ slightly above $T_{\rm c}$. Figures~\ref{Fig2}(a-c), \ref{Fig2}(d-e), \ref{Fig2}(f-h) show the Fermi surface, effective dispersion $\xi_{\gamma {\bm k}}^{\rm eff} = \varepsilon_{\gamma {\bm k}} + {\rm Re }\Sigma_{\gamma {\bm k}}(0) - \mu $, and the self-energy, respectively. 

In the noninteracting case, at half filling ($\delta\!=\!0$), the system behaves as a metal with electron and hole pockets centered around ${\bm k}=(0,0)$ and $(\pi,\pi)$, respectively.
These pockets are equal in size, and hole doping reduces (increases) the size of the electron (hole) pocket.

Electron correlations, however, significantly alter this picture. 
At $\delta=0$, the system transitions into an insulator due to {\it dynamical} band repulsion~\cite{PhysRevB.75.193103}. 
The frequency dependence of the real part of the self-energy plays a crucial role, pushing the antibonding band upward and the bonding band downward at low frequencies [Fig.~\ref{Fig2}(f)]. 
Upon doping, this {\it correlated} band insulator evolves into a metallic state.
Remarkably, in the underdoped regime ($\delta=0.02$), 
we do not see sharp spectral intensity for the electron pocket [Fig.~\ref{Fig2}(a)]. 
One might attribute this to dynamical band repulsion making the electron pocket incipient. 
However, the real part of the self-energy for the antibonding band bends downward at low Matsubara frequencies, counteracting the band repulsion [Fig.~\ref{Fig2}(f)]. 
If the imaginary part of the self-energy were neglected, the electron pocket would persist [red curve in Fig.~\ref{Fig2}(d)]. 
Instead, the enhancement of the imaginary part of the self-energy leads to a pseudogap in the antibonding band [Fig.~\ref{Fig2}(g)]. 
This suggests that the system tends to incorporate the antibonding band as an active component for SC, and the enhancement of self-energy would result from the formation of preformed pairs, as observed in the BEC regime of the attractive Hubbard model~\cite{Sakai_2015}. 
Interestingly, it is the hole pocket that becomes nearly incipient in this regime [blue curve in Fig.~\ref{Fig2}(d)] despite doping holes. 
If we neglect the imaginary part of the self-energy, the size of the electron pocket is unexpectedly larger than that of the hole pocket, in stark contrast to the noninteracting case.

In the overdoped regime ($\delta=0.12$), there are no such drastic correlation effects in the electronic structure. 
While the Fermi pockets are smaller than in the noninteracting case [Fig.~\ref{Fig2}(c)], this reduction can be explained by a weakly frequency-dependent, Hartree-like band shift [Figs.~\ref{Fig2}(e,f)]. 
Furthermore, the imaginary part of the self-energy shows Fermi-liquid-like behavior for both the bonding and antibonding bands [Figs.~\ref{Fig2}(g,h)].

\begin{figure}[tb]
\vspace{0.0cm}
\begin{center}
\includegraphics[width=0.49\textwidth]{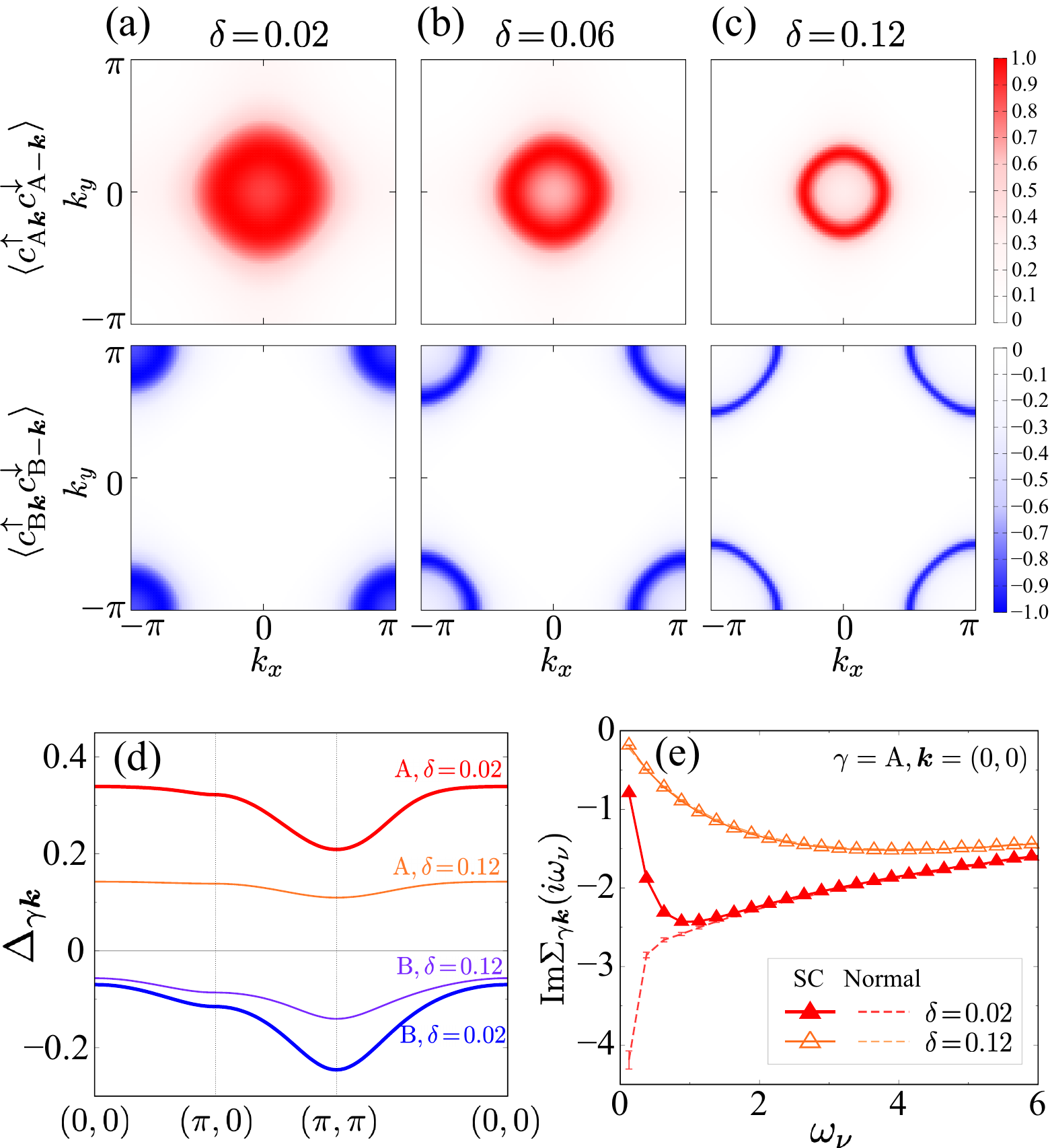}
\caption{
The results of SC state at $T=0.04$.
(a,b,c) The pairing amplitude $\langle c_{\gamma {\bm k}}^\uparrow  c_{\gamma -{\bm  k}}^\downarrow \rangle$ in momentum space (normalized by the maximum value). 
(d) The SC gap function $\Delta_{\gamma {\bm k}}$, which is estimated from $\Delta_{\gamma {\bm k}} = Z_{\gamma {\bm k}} S_{\gamma {\bm k}} (i\omega_0)$ ($Z_{\gamma {\bm k}}$ is the quasiparticle weight).
(e) Comparison of the normal part of the self-energy $\Sigma_{\gamma {\bm k}} (i \omega_\nu )$ between SC and normal states. 
}
\vspace{0cm}
\label{Fig3}
\end{center}
\end{figure}

\paragraph{Properties of SC state.}
Figures~\ref{Fig3}(a-c), \ref{Fig3}(d), \ref{Fig3}(e) illustrate the momentum dependence of the pairing amplitude, gap function, and a comparison of the self-energy between the normal and SC states, respectively.
In the underdoped regime ($\delta\!=\!0.02$), the system exhibits a strong-coupling nature.
The gap function becomes large [Fig.~\ref{Fig3}(d)], reaching a magnitude comparable to the Fermi energy.
As a result, the pairing amplitude distribution in momentum space broadens substantially [Fig.~\ref{Fig3}(a)], indicating the formation of localized Cooper pairs in real space.
In this region, the divergence of the imaginary part of the self-energy for the antibonding band, characteristic of the pseudogap state, vanishes in the SC phase due to the emergence of coherence [Fig.~\ref{Fig3}(e)], resulting in a kinetic-energy gain.
This dissolution of Im$\Sigma$ in the SC phase is similar to the behavior in the BEC regime of the attractive Hubbard model~\cite{Sakai_2015}.
As doping increases, the gap size decreases, and the pairing amplitude distribution becomes more concentrated around the Fermi level. 
This evolution of the SC state, crossing over from a BEC-like regime to a Bardeen-Cooper-Schrieffer (BCS)-like regime~\cite{RevModPhys.96.025002}, aligns with the corresponding change of the normal state from a pseudogap regime to a conventional metallic regime.

The crossover from kinetic-energy-driven SC to potential-energy-driven SC as a function of doping is reminiscent of a similar behavior reported in previous cluster DMFT studies on the single-layer Hubbard model~\cite{Maier_2004,Gull_2012,Fratino_2016}.
These studies demonstrated kinetic-energy-driven SC in the underdoped region, where a pseudogap forms at temperatures far above $T_{\rm c}$.
In this region, strong short-range singlet correlations above $T_{\rm c}$ limit the potential energy gain across the transition, while the emergence of coherence leads to a reduction in kinetic energy.
The well-known dichotomy between antinodal and nodal regions is replaced by a dichotomy between the antibonding and bonding bands in this system.
It is highly suggestive and intriguing that high-$T_{\rm c}$ SC in both single-layer and bilayer Hubbard models shares the common feature of momentum-space dichotomy in the parent normal phases.
It is also interesting that the dichotomy here arises from the difference in self-energy between bonding and antibonding orbitals, making it closely related to orbital-selective physics in iron-based superconductors~\cite{Misawa_2012,M_Yi_2013,de_Medici_2014,Kim_2024}.

Finally, we examine the coherence length $\xi$ of the present $s_\pm$-wave SC.
For this purpose, we employ a recently developed method~\cite{npjQuantumMat.9.100}.
We use $2 \! \times \! 1\! \times \! 1$ cDMFT with an impurity problem consisting of two interlayer sites, and impose a finite-momentum pairing state by introducing an order parameter modulation of the Fulde-Ferrell type $\Psi_{\bm q}({\bm r}) = |\Psi_{\bm q}| e^{i {\bm q}\cdot {\bm r}}$~\cite{Fulde_1964}.
From a Ginzburg-Landau expansion of the free energy, one can derive that the criterion for the disappearance of $|\Psi_{\bm q}|$ is given by $\xi |{\bm q}| \sim 1$.
Thus, by analyzing how robust the SC is against shifts in the center-of-mass momentum of the Cooper pairs, we can estimate $\xi$.

\begin{figure}[tb]
\vspace{0.0cm}
\begin{center}
\includegraphics[width=0.49\textwidth]{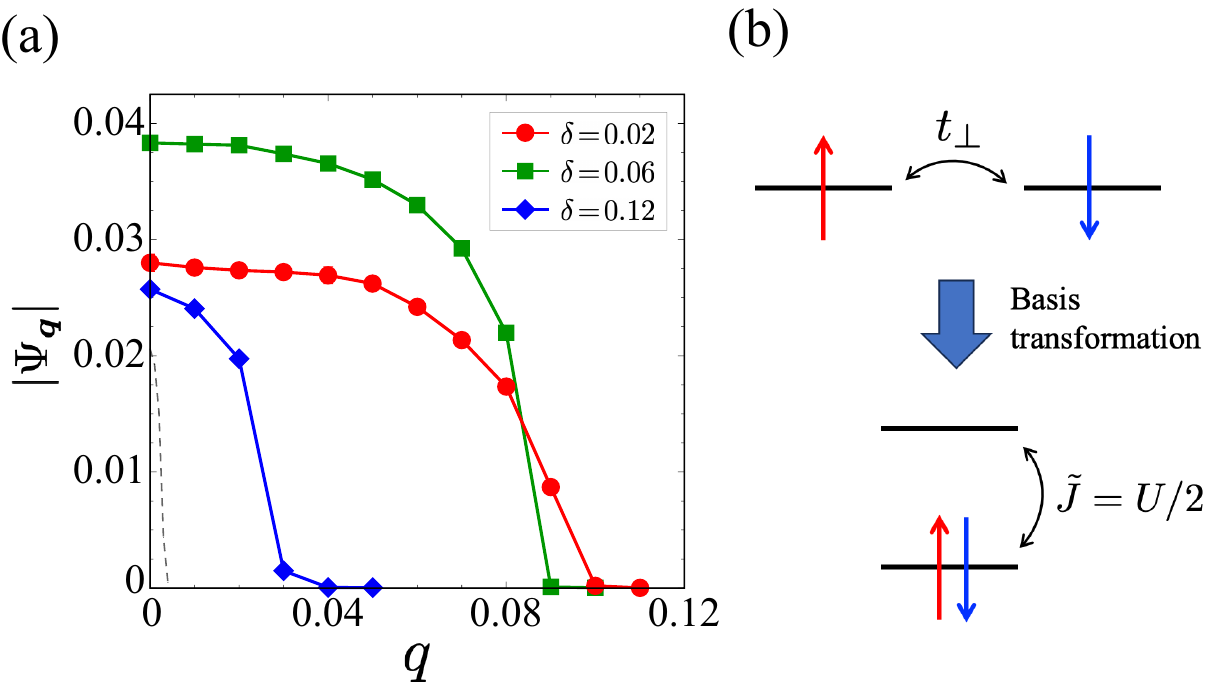}
\caption{
(a) Order parameter of finite-momentum pairing state at $T=0.04$. 
The dashed curve shows the result for the single-layer attractive Hubbard model ($U\!=\!-1$) on the square lattice at half filling. 
(b) Mapping from the bilayer Hubbard model to an effective two-orbital model.
}
\vspace{0cm}
\label{Fig4}
\end{center}
\end{figure}

Figure~\ref{Fig4}(a) displays the order parameter $\Psi_{\bm q} = \frac{1}{N_{\bm k}}\sum_{\bm k} \langle c_{1 {\bm k}+{\bm q}/2}^\uparrow c_{2 -{\bm k}+{\bm q}/2}^\downarrow \rangle$ for ${\bm q} =  q(2\pi,0)$, where the phase of the interlayer pairing varies along the $x$-direction.
The results indicate a very short coherence length, on the order of a few in-plane lattice constants, consistent with the broad distribution of the pairing amplitude in momentum space [Figs.~\ref{Fig3}(a-c)].
The localized nature of Cooper pairs is also evident when compared to the behavior observed in the BCS regime of the attractive Hubbard model [dashed curve in Fig.~\ref{Fig4}(a)].

By applying the basis transformation to the bonding-antibonding basis, the bilayer Hubbard model can be mapped onto a two-orbital model~\footnote{Please refer to Ref.~\cite{Ferrero_2009,Shinaoka_2015} and Supplemental Materials at http://*** for the detailed derivation for the basis transformation.} with an effective pair-hopping term of $U/2$ [Fig.~\ref{Fig4}(b)].
In the context of fulleride superconductors~\cite{Nomura_2016}, it has been reported that a strong pair-hopping term significantly shortens the coherence length~\cite{npjQuantumMat.9.100}. 
It would be interesting to revisit the extremely short coherence length observed in this study from the viewpoint of effective multi-orbital physics.

\paragraph{Discussion.}

The bilayer Hubbard model can be regarded as a minimal model for $d_{3z^2-r^2}$ bands in La$_3$Ni$_2$O$_7$. Since this model does not include $d_{x^2-y^2}$ degrees of freedom, care must be taken when discussing its relevance to La$_3$Ni$_2$O$_7$. However, if the $d_{3z^2-r^2}$ orbital plays an active role in La$_3$Ni$_2$O$_7$, the physics discussed here may provide some insights into the nature of high-$T_{\rm c}$ SC in La$_3$Ni$_2$O$_7$.
In particular, our finding of an extremely short coherence length $\xi$ of just a few lattice constants is highly suggestive. 
Given that the in-plane Ni-Ni distance in La$_3$Ni$_2$O$_7$ is approximately 3.8 \AA, our results indicate a coherence length on the order of 1 nm. This naturally explains the extraordinarily high critical field, which reaches $\sim 100$ T~\cite{Nature.621.493,arxiv.2501.14584}.
Furthermore, it is fascinating that nickelates provide a unique platform where both single-layer physics (doped $R$NiO$_2$ with $R$=La, Pr, Nd, and Sm-Eu~\cite{Li_2019,Kitatani_2020,Nomura_2022,Chow_2025}) and bilayer physics (La$_3$Ni$_2$O$_7$) can be explored simultaneously. 
This dual perspective will give a deeper understanding of high-$T_{\rm c}$ SC.

\paragraph{Conclusion.}
We have explored the nature of $s_\pm$-wave SC in the bilayer Hubbard model. Unlike $d$-wave superconductivity, which emerges from doping the Mott insulator, the SC state here arises from doping a correlated band insulator.
Nevertheless, a strong momentum-space dichotomy characterized by pseudogap behavior and a crossover between kinetic-energy-driven and potential-energy-driven SC appear as common and highly suggestive features. Additionally, the connection to another high-$T_{\rm c}$ system, fullerides, is also intriguing, as the strong effective pair-hopping term may play a crucial role in realizing SC with an extremely short coherence length $\xi$.
Clarifying the potential relevance of this physics to La$_3$Ni$_2$O$_7$ remains a fundamentally important open question.

\paragraph{Acknowledgements.}
We acknowledge the financial support by Grants-in-Aid for Scientific Research (JSPS KAKENHI) [Grant Nos. JP23H04869 (YN), JP23H04519 (YN), JP23K03307 (YN), JP23H03817 (MK), JP24K17014 (MK),  JP23H04528 (SS), JP22H00110 (RA), JP24H00190 (RA), and JP25H01252 (RA)] and MEXT as ``Program for Promoting Researches on the Supercomputer Fugaku'' (Project ID: JPMXP1020230411). This work was supported by the RIKEN TRIP initiative (RIKEN Quantum, Advanced General Intelligence for Science Program, Many-body Electron Systems).
Part of the calculations were performed at Supercomputer Center, ISSP, University of Tokyo.

\paragraph{Data Availability.}

The data that support the findings of this study are available in Zenodo~\cite{nomura_2025_15713498}. 

\bibliography{main,bilayer}

\clearpage

\appendix

\setcounter{section}{0}
\setcounter{equation}{0}
\setcounter{figure}{0}
\setcounter{table}{0}
\setcounter{page}{1}
\renewcommand{\theequation}{S\arabic{equation}}
\renewcommand{\thefigure}{S\arabic{figure}}
\renewcommand{\thetable}{S\arabic{table}}
\renewcommand{\thesection}{S\arabic{section}}

\begin{center}
    {\bf Supplemental Materials for \\ ``Strong-coupling high-$T_{\rm c}$ superconductivity in doped correlated band insulators'' \\ } 
    \vspace{0.2cm}
    {\bf \small Yusuke Nomura, Motoharu Kitatani, Shiro Sakai, and Ryotaro Arita}
 \end{center}

\subsection{Mapping from the bilayer Hubbard model to an effective two-orbital model}

The basis transformation to bonding [$c^{\sigma}_{{\rm B} i}= \frac{1}{\sqrt{2}} ( c^{\sigma}_{1 i} + c^{\sigma}_{2 i})$] and antibonding [$c^{\sigma}_{{\rm A} i}= \frac{1}{\sqrt{2}} ( c^{\sigma}_{1 i} - c^{\sigma}_{2 i})$] basis converts the bilayer Hubbard model into an effective two-orbital model. 
The one-body part of the Hamiltonian is transformed to 
\begin{eqnarray}
 \mathcal{H}_0 = \sum_{\gamma={\rm A,B}} \sum_{{\bm k},\sigma} \varepsilon_{\gamma {\bm k}} c^{\sigma \dagger}_{\gamma {\bm k}} c^{\sigma}_{\gamma {\bm k}}, 
\end{eqnarray}
where the energy dispersions are given by $\varepsilon_{{\rm A} {\bm k}}=t_\perp -2t(\cos k_x + \cos k_y)$ for the antibonding orbital ($\gamma={\rm A}$) and $\varepsilon_{{\rm B} {\bm k}}=-t_\perp -2t(\cos k_x + \cos k_y)$ for the bonding orbital ($\gamma={\rm B}$).
The interaction part of the effective two-orbital Hamiltonian is given by
\begin{eqnarray}
 \mathcal{H}_{\rm int}  \! 
&=&   \!  
 \tilde{U} \sum_{i,\gamma} n^{\uparrow}_{\gamma i}  n^{\downarrow}_{\gamma i } 
 +   
\tilde{U}'    \sum_{i,\sigma} 
 n^{\sigma}_{{\rm A} i}  n^{\! -\sigma} _{{\rm B} i}     +   
(\tilde{U}' \! -\! \tilde{J})  \sum_{i,\sigma}   
n^{\sigma}_{{\rm A} i}  n^{\sigma}_{{\rm B} i} 
 \nonumber \\ 
\!   &+&  \!  
 \tilde{J} \!  \sum_{i, \gamma \neq \gamma' }  
c^{\uparrow \dagger}_{\gamma i} 
c^{\uparrow}_{\gamma' i} 
c^{\downarrow \dagger}_{\gamma' i}
c^{\downarrow}_{\gamma i}
+   
\tilde{J}  \!  \sum_{i, \gamma \neq \gamma'} 
c^{\uparrow \dagger }_{\gamma i} 
c^{\uparrow}_{\gamma' i} 
c^{\downarrow \dagger}_{\gamma i}
c^{\downarrow}_{\gamma' i}, 
\end{eqnarray}
where the interaction parameters are defined as $\tilde{U}=\tilde{U}'=\tilde{J}=U/2$.
In particular, the last term corresponds to the effective pair-hopping term depicted in Fig.~\ref{Fig4}(b), which provides a direct pairing interaction.
Since the sign of this pairing interaction is positive, the phase of the pairing exhibits a sign change between the bonding and antibonding bands, leading to the pairing symmetry of $s_{\pm}$-wave.

\end{document}